\documentclass[journal=aamick,manuscript=article]{amsart}
\usepackage[version=3]{mhchem} 
\usepackage[T1]{fontenc}  
\usepackage[super]{natbib}
\usepackage{physics}
\calclayout
\usepackage[foot]{amsaddr}
\usepackage{subfig}
\usepackage{mathtools}
\usepackage{amssymb}
\usepackage{gensymb}
\usepackage{graphicx}
\usepackage{filecontents}
\usepackage{bm}
\usepackage{xr}
\newcommand{\beginsupplement}{%
        \setcounter{table}{0}
        \renewcommand{\thetable}{S\arabic{table}}%
        \setcounter{figure}{0}
        \renewcommand{\thefigure}{S\arabic{figure}}%
        \setcounter{equation}{0}
        \renewcommand{\theequation}{S\arabic{equation}}
     }

\title{The thermo-wetting instability driving Leidenfrost film collapse}

\author{Tom Y. Zhao$^\dagger$}

\author{Neelesh A. Patankar$^{\dagger,*}$}
\address[$^\dagger$]{Northwestern University, Department of Mechanical Engineering: 2145 Sheridan Road, Evanston, Illinois 60208, USA}
\address[$^*$]{E-mail: n-patankar@northwestern.edu}

\begin{document}

\maketitle
\begin{abstract}
\normalsize Above a critical temperature known as the Leidenfrost point (LFP), a heated surface can suspend a liquid droplet above a film of its own vapor. The insulating vapor film can be highly detrimental in metallurgical quenching and thermal control of electronic devices, but may also be harnessed to reduce drag and generate power. Manipulation of the LFP has occurred mostly through experiment, giving rise to a variety of semi-empirical models that account for the Rayleigh-Taylor instability, nucleation rates, and superheat limits. However, a truly comprehensive model has been difficult given that the LFP varies dramatically for different fluids and is affected by system pressure, surface roughness and liquid wettability. Here, we investigate the vapor film instability for small length scales that ultimately sets the collapse condition at the Leidenfrost point. From a linear stability analysis, it is shown that the main film stabilizing mechanisms are the liquid-vapor surface tension driven transport of vapor mass and evaporation at the liquid-vapor interface. Meanwhile, van der Waals interaction between the bulk liquid and the solid substrate across the vapor phase drives film collapse. This physical insight into vapor film dynamics allows us to derive an ab-initio, mathematical expression for the Leidenfrost point of a fluid. The expression captures the experimental data on the LFP for different fluids under various surface wettabilities and ambient pressures. For fluids that wet the surface (small intrinsic contact angle), the expression can be simplified to a single, dimensionless number that encapsulates the wetting instability governing the LFP.
\end{abstract}
\section{Introduction}

As a surface is superheated above the boiling point of an adjacent fluid, vapor bubbles nucleate and grow. The boiling behavior of the liquid phase undergoes a fundamental change at a critical temperature known as the Leidenfrost point. Beyond this point, a film of insulating vapor forms between the liquid and the surface that suppresses heat transfer from the solid material. This heat flux reduction can be highly detrimental in the quenching of metal alloys by extending cooling rates and precluding the desired increase in strength and hardness \cite{2017Quench}. Alternatively, film boiling may be used to promote drag reduction as well as enable power generation through self-propulsion \cite{2011Drag, 2012Friction, 2015heat}. Thus, modulation of the LFP through fluid choice, surface texture and chemistry for the specific application is crucial \cite{2013Varanasi}. 

On a fundamental level, the physical mechanism responsible for the LFP is still uncertain. Many theoretical frameworks have been used to characterize the Leidenfrost effect and estimate the LFP, including hydrodynamic instability \cite{1958Zuber, 1961Berenson}, superheat spinodal limits \cite{1992Carey, 1963Spiegler}, and the change of liquid wettability on the heated surface with temperature \cite{1973Adamson, 1980Segev}. A thermocapillary model has also been proposed that attributes the film instability to fluctuations at micron length scales; however the analysis posits that the thermocapillary effect is the dominant destabilizing term, which does not explain the significant change in LFP on surfaces with different wettabilities \cite{2018Thermocap}. For example, the LFP of water can vary from $300\degree \text{C}$ for hydrophilic surfaces to $145\degree \text{C}$ on hydrophobic surfaces  \cite{2013Nanotubes, 2013Kruse}.

In this work, we introduce a stability analysis of the vapor film at the nanoscale regime. The dominant destabilizing term arises from the van der Waals interaction between the bulk liquid and the substrate across a thin vapor layer. On the other hand, liquid-vapor surface tension driven transport of vapor and evaporation at the two phase interface stabilize the film. The competition between these mechanisms gives rise to a comprehensive description of the LFP as a function of both fluid and solid properties. For fluids that wet the surface, such that the intrinsic contact angle is small, a single dimensionless number (eqn. \ref{eqn. pi16}) can be derived that encapsulates the instability determining the Leidenfrost point. 

It is noted that the literature has proposed different names for the critical temperature associated with a droplet levitating on a heated plate (Leidenfrost point) versus the critical temperature for vapor film formation in pool boiling (minimum film temperature). The Leidenfrost point has been shown to be equivalent to the minimum film boiling temperature for saturated liquids on isothermal surfaces \cite{1973minleid}. In this work, the term LFP will be used for both cases as a matter of convenience, with the understanding that no undercooling is applied to the liquid phase for pool boiling scenarios unless explicitly stated.  

\section{Film Instability}
There are many approaches to examine the stability of a vapor film adjacent to a superheated wall in two dimensions. Models have been developed with a base solution imposing static equilibrium, where the interface is at the saturation temperature corresponding to the imposed, far field liquid pressure \cite{2018Thermocap, 1992flat}. Here, we consider the thickness of the vapor film to be in dynamic equilibrium, as in a vertical plate configuration \cite{1993CHT} or vapor under a droplet \cite{2003Quere}. This appears to be a more general analysis since under experimental settings, droplet levitation occurs over a film that is continuously replenished by evaporation and depleted through escape of the buoyant vapor phase. Similarly in a horizontal setup for pool boiling, bubbles pinch off the film, necessitating a nonzero rate of evaporation to sustain a constant mean film thickness \cite{1993CHT}.  
\begin{figure}[]
\centering
\includegraphics[width=5cm]{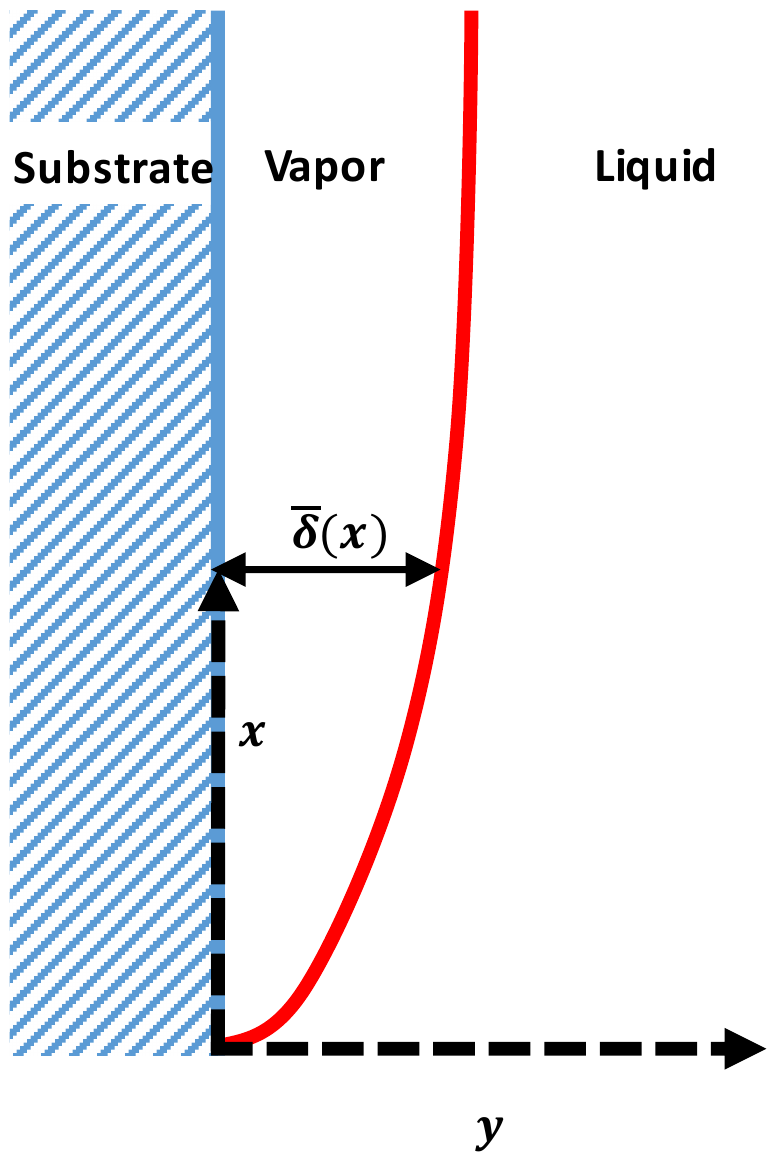}
\caption{Film boiling on a vertical plate, with the coordinate system delineated. The film thickness is denoted by $\bar{\delta}(x)$}
\label{fig: nicegraphic}
\end{figure}

Fig. \ref{fig: nicegraphic} shows a chosen ``model'' problem of film boiling on a vertical plate \cite{1993CHT, 1992Carey}. The vapor forms a laminar layer at the wall, with evaporation at the liquid interface sustaining the buoyant transport of vapor mass away from the base of the plate. The surrounding liquid is saturated and motionless with its properties fixed at the saturation temperature. The properties of vapor are assumed to be constant at the superheated wall temperature.
\subsection{Governing Equations}
The mass, momentum and energy conservation equations in the vapor domain are given by the following boundary-layer equations:
\begin{equation}
\pdv{\bar{u}_{\scriptscriptstyle V}}{x}+\pdv{\bar{v}_{\scriptscriptstyle V}}{y}=0
\label{eqn: mass conservation}
\end{equation}
\begin{equation}
\bar{u}_{\scriptscriptstyle V}\pdv{\bar{u}_{\scriptscriptstyle V}}{x}+\bar{v}_{\scriptscriptstyle V}\pdv{\bar{u}_{\scriptscriptstyle V}}{y}=-\frac{1}{\rho_{\scriptscriptstyle V}}\pdv{\bar{\Phi}}{x}+\frac{\Delta \rho g}{\rho_{\scriptscriptstyle V}}+\frac{\mu_{\scriptscriptstyle V}}{\rho_{\scriptscriptstyle V}}\pdv[2]{\bar{u}_{\scriptscriptstyle V}}{y}
\label{eqn: mom conservation}
\end{equation}
\begin{equation}
\pdv{\bar{\Theta}}{t}+\bar{u}_{\scriptscriptstyle V}\pdv{\bar{\Theta}_{\scriptscriptstyle V}}{x}+\bar{v}_{\scriptscriptstyle V}\pdv{\bar{\Theta}_{\scriptscriptstyle V}}{y}=\frac{k_{\scriptscriptstyle V}}{\rho_{\scriptscriptstyle V}c_{p,\scriptscriptstyle V}}\pdv[2]{\bar{\Theta}_{\scriptscriptstyle V}}{y}
\label{eqn: energy conservation}
\end{equation}
where the parameters $\mu$, $\rho$, $k$, g and $c_{p}$ represent the the dynamic viscosity, density, thermal conductivity, gravitational acceleration and specific heat of the fluid, respectively. The term $\Delta\rho=\rho_{\scriptscriptstyle L}-\rho_{\scriptscriptstyle V}$ represents the density difference, the subscripts $L$ and $V$ denote the liquid and vapor field, and the temperature has been normalized as $\bar{\Theta}=\frac{\bar{T}-T_s}{T_w-T_s}$, the difference between the temperature field and the saturation temperature $T_s$ at the interface over the difference between the wall temperature $T_w$ and $T_s$. Note that we take the liquid phase to be motionless due to its much greater viscosity relative to the vapor ($u_{\scriptscriptstyle L}(x,t)=0$), as well as isothermal at saturation temperature due to its larger thermal diffusivity ($\Theta_{\scriptscriptstyle L}=0$). This simplification allows for an analytical solution for the base state per Burmeister \cite{1993CHT} .

The generalized pressure term $\bar{\Phi}$ takes into account both the fluid pressure arising due to surface tension as well as due to van der Waals interactions. To first order in the base solution, these terms are negligible since the liquid-vapor interface is assumed to be locally parallel to the wall \cite{1984BLStab, 1996LinStab}; this implies $\bar{\Phi}\approx0+\Phi'$, where the overbar variables represent the general solution, the unbarred variables denote the base solution, and the primed variables give the perturbed solution. Additionally, the momentum and temperature equations are modeled as steady in the base solution and only exhibit a time varying term in the linearized equations for the perturbations. 

The boundary conditions at the superheated wall and the liquid-vapor interface at $\bar{\delta}(x)$ are given by:
 \begin{equation}
\text{at} \; y=0, \quad \bar{u}_{\scriptscriptstyle V}=\bar{v}_{\scriptscriptstyle V}=0\;, \quad \bar{\Theta}_{\scriptscriptstyle V}=1
\label{eqn: wallbcs}
\end{equation}
 \begin{equation}
\text{at}\; y=\bar{\delta}, \quad \bar{u}_{\scriptscriptstyle V}+\bar{v}_{\scriptscriptstyle V}\dv{\bar{\delta}}{x}=0\;, \quad \bar{\Theta}_{\scriptscriptstyle V}=0
\label{eqn: intbcs}
\end{equation}

Eqn. \ref{eqn: wallbcs} enforces the temperature and no interfacial slip at the impermeable wall, while eqn. \ref{eqn: intbcs} ensures that the tangential component of velocity is continuous and that the temperature is at saturation along the liquid-vapor interface. As with Burmeister, we neglect the "blowing" of vapor toward the plate by assuming $v(x,t)\approx0$, such that the preceding set of boundary conditions are sufficient to fully specify the problem. Otherwise, the normal component of velocity would also need to be fixed at the interface; this leads to a cubic rather than a parabolic estimate to the velocity field.

The vapor generated due to phase change at this interface is balanced by the streamwise rate of change of the vapor flow in the film and the growth of the film thickness in time:
\begin{equation}
\rho_{\scriptscriptstyle V}\pdv{\bar{\delta}}{t}+\pdv{}{x}\int_0^{\bar{\delta}}(\rho_{\scriptscriptstyle V}\bar{u}_{\scriptscriptstyle V}dy)=-\frac{k_{\scriptscriptstyle V}}{h_{\scriptscriptstyle LV}}\pdv{\bar{\Theta}_{\scriptscriptstyle V}}{y}|_{y=\bar{\delta}}
\label{eqn: basedelta}
\end{equation}
where $h_{\scriptscriptstyle LV}$ is the latent heat of vaporization. In the base state, the time variation of the film thickness is taken to be negligible, $\dv{\bar{\delta}}{t}\approx0+\dv{\delta'}{t}$.
\subsection{Base Flow}
Along with the boundary conditions (eqn. \ref{eqn: wallbcs} and \ref{eqn: intbcs}), the base flow equations are:
\begin{equation}
\pdv{u_{\scriptscriptstyle V}}{x}+\pdv{v_{\scriptscriptstyle V}}{y}=0
\label{eqn: mass conservation}
\end{equation}
\begin{equation}
u_{\scriptscriptstyle V}\pdv{u_{\scriptscriptstyle V}}{x}+v_{\scriptscriptstyle V}\pdv{u_{\scriptscriptstyle V}}{y}=\frac{\Delta \rho g}{\rho_{\scriptscriptstyle V}}+\frac{\mu_{\scriptscriptstyle V}}{\rho_{\scriptscriptstyle V}}\pdv[2]{u_{\scriptscriptstyle V}}{y}
\label{eqn: mom conservation}
\end{equation}
\begin{equation}
u_{\scriptscriptstyle V}\pdv{\Theta_{\scriptscriptstyle V}}{x}+v_{\scriptscriptstyle V}\pdv{\Theta_{\scriptscriptstyle V}}{y}=\frac{k_{\scriptscriptstyle V}}{\rho_{\scriptscriptstyle V}c_{p,\scriptscriptstyle V}}\pdv[2]{\Theta_{\scriptscriptstyle V}}{y}
\label{eqn: energy conservation}
\end{equation}

The steady, developed solution can be found approximately by using an integral expansion method, which is described in full detail by Burmeister \cite{1993CHT}. After introducing the normalized variable $\eta=y/\delta$, we can determine the velocity ($u_{\scriptscriptstyle V}$) and temperature field ($\Theta_{\scriptscriptstyle V}$) in the base solution:
\begin{equation}
u_{\scriptscriptstyle V} = \frac{\Delta\rho g \delta^2}{2\mu_{\scriptscriptstyle V}}(\eta-\eta^2)
\end{equation}
\begin{equation}
\Theta_{\scriptscriptstyle V}=\frac{T_{\scriptscriptstyle V}-T_s}{T_w-T_s}=1+(-1-\frac{1-c}{2})\eta+\frac{1-c}{2}\eta^3
\end{equation}
Note that the velocity profile is locally parabolic due to the buoyancy force, while the velocity variation in the streamwise direction under mass conservation is encapsulated in the $\delta(x)^2$ dependence and occurs on a much larger length scale. The value of $c$ can be found by solving the quadratic expression:
\begin{equation}
\frac{1}{3}\frac{c_{p,\scriptscriptstyle V}\Delta T}{h_{\scriptscriptstyle LV}}c\left(1-\frac{3}{10}(1-c)\right)=1-c
\end{equation}
and $\Delta T=T_w-T_s$. For typical Jakob numbers around $Ja=\frac{c_{p,\scriptscriptstyle V}\Delta T}{h_{\scriptscriptstyle LV}}=\frac{1}{10}$, $c$ can be found from a simplified linear equation $c=1-\frac{1}{3}\frac{c_{p,\scriptscriptstyle V}\Delta T}{h_{\scriptscriptstyle LV}}$. Using this approximation, the film thickness $\delta$ is described by:
\begin{equation}
\delta = 2\left(1-\frac{1}{3}\frac{c_{p,\scriptscriptstyle V}\Delta T}{h_{\scriptscriptstyle LV}}\right)^{1/4}\left(\frac{x\Delta T\mu_{\scriptscriptstyle V}k_{\scriptscriptstyle V}}{\rho_{\scriptscriptstyle V}h_{\scriptscriptstyle LV}g\Delta\rho}\right)^{1/4}
\end{equation}
Note that to first order, the velocity and temperature fields as well as the film thickness are steady.
\subsection{Linearized Equations}
The base solutions for the velocity, temperature and film thickness are perturbed, yielding the following linearized equations:
\begin{equation}
\pdv{u'}{x}+\pdv{v'}{y}=0
\label{eqn: mass conservation}
\end{equation}
\begin{equation}
u'\pdv{u}{x}+u\pdv{u'}{x}+v'\pdv{u}{y}+v\pdv{u'}{y}=-\frac{1}{\rho_{\scriptscriptstyle V}}\pdv{\Phi'}{x}+\frac{\mu_{\scriptscriptstyle V}}{\rho_{\scriptscriptstyle V}}\pdv[2]{u'}{y}
\label{eqn: mom conservation}
\end{equation}
\begin{equation}
\pdv{\Theta'}{t}+u'\pdv{\Theta}{x}+u\pdv{\Theta'}{x}+v'\pdv{\Theta}{y}+v\pdv{\Theta'}{y}=\frac{k_{\scriptscriptstyle V}}{\rho_{\scriptscriptstyle V}c_{p,\scriptscriptstyle V}}\pdv[2]{\Theta'}{y}
\label{eqn: energy conservation}
\end{equation}
The boundary conditions at the wall are:
\begin{equation}
\text{at}\; y=0 \quad u'_{\scriptscriptstyle V}=v'_{\scriptscriptstyle V}=0\;, \quad \Theta'_{\scriptscriptstyle V}=0
\label{eqn: wallbcsper}
\end{equation}
At the perturbed interface location $\delta+\delta'$, the tangential velocity and temperature conditions (eqn. \ref{eqn: intbcs}) after applying the locally parallel approximation $\dv{\delta}{x}\approx0$ give:
\begin{equation}
u_{\scriptscriptstyle V}'|_{\eta=1}+\frac{\delta'}{\delta}\pdv{u}{\eta}|_{\eta=1}=0\;, \quad \Theta'_{\scriptscriptstyle V}|_{\eta=1}+\frac{\delta'}{\delta}\pdv{\Theta_{\scriptscriptstyle V}}{\eta}|_{\eta=1}=0
\label{eqn: intbcsper}
\end{equation}
The phase change equation at the interface (eqn. \ref{eqn: basedelta}) is linearized as:
\begin{equation}
\rho_{\scriptscriptstyle V}\pdv{\delta'}{t}+\pdv{}{x}\int_0^\delta\rho_{\scriptscriptstyle V}u'_{\scriptscriptstyle V}dy+\pdv{}{x}\int_0^{\delta+\delta'}\rho_{\scriptscriptstyle V}u_{\scriptscriptstyle V}dy=-\frac{k\Delta T}{h_{\scriptscriptstyle LV}}\left(\delta'\pdv[2]{\Theta}{y}|_\delta+\pdv{\Theta'}{y}|_{\delta}\right)
\label{eqn: deltalin}
\end{equation}
Analogous to the base solution, the perturbed velocity is expanded in powers of $\eta$. 
\begin{equation}
u'_{\scriptscriptstyle V}=a'_0+a_1'\eta+a_2'\eta^2
\label{eqn: upexan}
\end{equation}
\begin{equation}
\text{at} \; \eta=0 \quad u_{\scriptscriptstyle V}'=0 \rightarrow a_0'=0
\end{equation}
The terms $a_1'$ and $a_2'$ can be found as functions of the fluid properties and the generalized pressure gradient $\pdv{\Phi'}{x}=\pdv{p'}{x}+\pdv{\phi'}{x}$ from the momentum equation eqn. \ref{eqn: mom conservation} evaluated at the wall ($\eta=0$) and the tangential velocity condition (eqn. \ref{eqn: intbcsper}).
\begin{equation}
a_1'=\left(\frac{\Delta\rho g\delta\delta'}{2\mu_{\scriptscriptstyle V}}-\frac{\pdv{\Phi'}{x}\delta^2}{2\mu_{\scriptscriptstyle V}}\right)
\end{equation}
\begin{equation}
a_2'=\frac{\pdv{\Phi'}{x}\delta^2}{2\mu_{\scriptscriptstyle V}}
\end{equation}
The pressure gradient arises from the liquid-vapor surface tension $\sigma_{\scriptscriptstyle LV}$ at the two phase interface due to capillary pressure induced by local nonzero curvature:
\begin{equation}
\pdv{p'}{x}=-\sigma_{\scriptscriptstyle LV}\dv[3]{\delta'}{x}
\end{equation}
This implies that positive curvature corresponds to the center of curvature lying in the vapor domain, such that the vapor bulges into the liquid. Here, we also introduce the disjoining pressure term $\bar{\phi}$, which describes the van der Waals interaction between the fluid and the substrate:
\begin{equation}
\bar{\phi}=\frac{A}{6\pi\bar{\delta}^3}
\end{equation}
The streamwise derivative of this term is negligible in the base state under the locally parallel interface approximation. The Hamaker constant $A$ is typically positive, denoting attractive interactions between dipoles \cite{2005Apos}. The perturbed component is:
\begin{equation}
\pdv{\phi'}{x}=-\frac{A}{2\pi\delta^4}\dv{\delta'}{x}
\end{equation}
This gives an expression for the perturbed, generalized pressure term evaluated at the liquid-vapor interface.
\begin{equation}
\pdv{\Phi'}{x}=\pdv{p'}{x}+\pdv{\phi'}{x}=-\sigma_{\scriptscriptstyle LV}\dv[3]{\delta'}{x}-\frac{A}{2\pi\delta^4}\dv{\delta'}{x}
\label{eqn: genP}
\end{equation}
Next, the expanded perturbed temperature is:
\begin{equation}
\Theta_{\scriptscriptstyle V}'=b_0'+b_1'\eta+b_2'\eta^2+b_3'\eta^3
\label{eqn: thetapexan}
\end{equation}
From the wall boundary condition (eqn. \ref{eqn: wallbcsper}) and energy conservation equation (eqn. \ref{eqn: energy conservation}) at $\eta=0$, we find that $b_0'=b_2'=0$. Similarly, the temperature condition (eqn. \ref{eqn: intbcsper}) and energy conservation (eqn. \ref{eqn: energy conservation}) at the interface $\eta=1$ leads to an expression for $b_1'$ in terms of $\delta'$:

\begin{multline}
\frac{\delta}{4}\pdv{b_1'}{t}+\frac{1}{4}(1-2b_3)\pdv{\delta'}{t}+(\frac{1}{6}-\frac{2b_3}{15})\frac{\Delta\rho g\delta^2}{2\mu_{\scriptscriptstyle V}}\dv{\delta'}{x}+(\frac{1}{12}-\frac{b_3}{20})(\frac{\delta^3}{2\mu_{\scriptscriptstyle V}})\left(\sigma_{\scriptscriptstyle LV}\dv[4]{\delta'}{x}+\frac{A}{2\pi\delta^4}\dv[2]{\delta'}{x}\right)\\+\frac{a_1}{30}(1-2b_3)\dv{\delta'}{x}+\frac{a_1\delta}{20}\dv{b_1'}{x}=\frac{3k_{\scriptscriptstyle V}}{\rho_{\scriptscriptstyle V}c_{p,\scriptscriptstyle V}\delta^2}\delta'-\frac{3k_{\scriptscriptstyle V}}{\rho_{\scriptscriptstyle V}c_{p,\scriptscriptstyle V}\delta}b_1'
\label{eqn: tempeqnperturb}
\end{multline}

The time evolution for the perturbed $\delta'$ as a function of $b'_1$ follows from the linearized phase change expression (eqn. \ref{eqn: deltalin}) and the expressions for $u_{\scriptscriptstyle V}'$ and $\Theta_{\scriptscriptstyle V}'$ (eqn. \ref{eqn: upexan} and \ref{eqn: thetapexan}): 
\begin{equation}
\rho_{\scriptscriptstyle V}\pdv{\delta'}{t}+\frac{\rho_{\scriptscriptstyle V}\Delta\rho g\delta^2}{4\mu_{\scriptscriptstyle V}}\dv{\delta'}{x}+\frac{\rho_{\scriptscriptstyle V}\delta^3\sigma_{\scriptscriptstyle LV}}{12\mu_{\scriptscriptstyle V}}\dv[4]{\delta'}{x}+\frac{\rho_{\scriptscriptstyle V}A}{24\pi\mu_{\scriptscriptstyle V}\delta}\dv[2]{\delta'}{x}+\frac{3k_{\scriptscriptstyle V}\Delta T}{h_{\scriptscriptstyle LV}\delta^2}\delta'-\frac{2k_{\scriptscriptstyle V}\Delta T}{h_{\scriptscriptstyle LV}\delta}b_1'=0
\label{eqn: delteqnperturb}
\end{equation}

\begin{figure*}[]
\centering
\includegraphics[width=17cm]{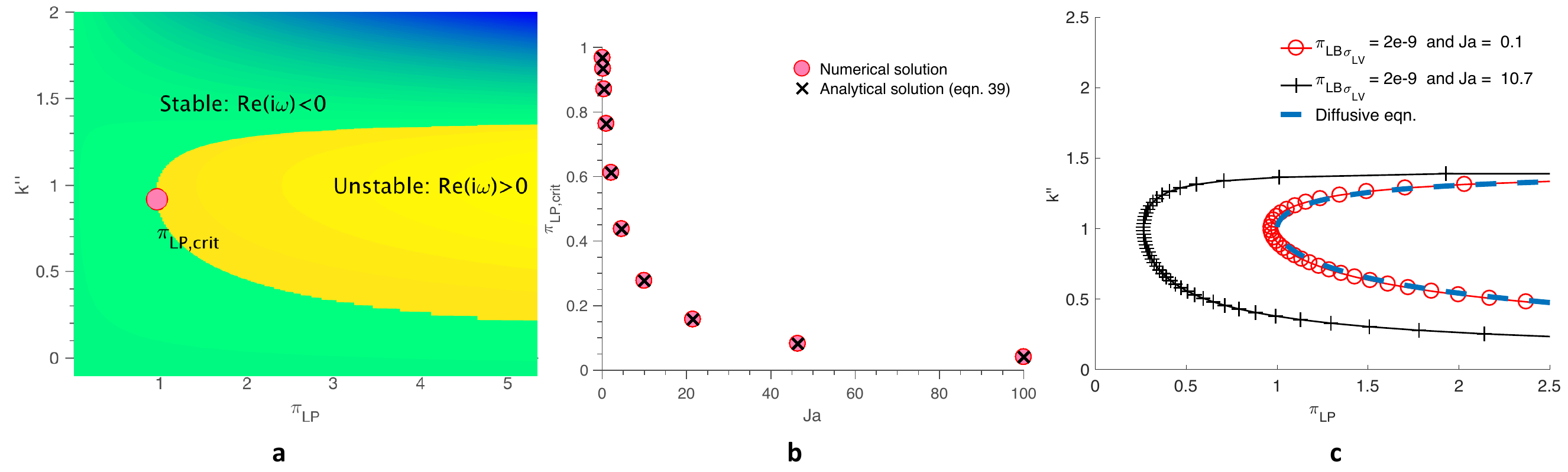}
\caption{\textbf{The stability of the perturbed solution.}\\ \textbf{a}. The variation of Re($i\omega$) with $\pi_{LP}$ and the dimensionless wavenumber $k''$ for $\Pi_{LB\sigma_{\scriptscriptstyle LV}}=2e-9$ and $Ja=0.1$. A critical $\pi_{LP,crit}$ can be defined, for which lower values lead to unconditional stability, and greater values allow the coexistence of stable and unstable zones. \textbf{b}. The Jakob number vs the critical $\pi_{LP, crit}$, as predicted by the numerical solution to the full stability solution (eqn. \ref{eqn: fulleqnnondim}) and by taking $\pi_{LP,crit}$ to occur at $k''=1$ (eqn: \ref{eqn: picritfull}). The agreement is excellent. \textbf{c} The marginal stability curves are calculated by taking the locus of points where the real value of $i\omega$ changes sign. The diffusive expression (eqn. \ref{eqn: diffeqnnondim}) is a good approximation for small Jakob numbers.}
\label{fig: discrimja}
\end{figure*}

The perturbation equations \ref{eqn: tempeqnperturb} and \ref{eqn: delteqnperturb} give two homogeneous conditions for $\delta'$ and $b_1'$. The perturbations can now be expressed in terms of normal modes:
\begin{equation}
\delta'=\delta_a' \text{exp}(i(kx+\omega t))
\end{equation}
\begin{equation}
b_1'=b_{1a}' \text{exp}(i(kx+\omega t))
\end{equation}
To avoid introducing new notation, we will represent the amplitudes without subscripts $\delta_a'\rightarrow\delta'$ and $b_{1a}'\rightarrow b_1'$. Here, $k$ is the wave number and $\omega$ the time rate of growth of the perturbation. We combine eqn. \ref{eqn: tempeqnperturb} and \ref{eqn: delteqnperturb} to obtain a single equation with the coefficient $\delta'$. To simplify the representation, we introduce the following dimensionless parameters:
\begin{equation}
\pi_{LP}=\frac{3A^2h_{\scriptscriptstyle LV}\rho_{\scriptscriptstyle V}}{(24\pi)^2\delta^3k_{\scriptscriptstyle V}\mu_{\scriptscriptstyle V}\Delta T\sigma_{\scriptscriptstyle LV}}
\end{equation} 
\begin{equation}
\pi_{LB\sigma_{\scriptscriptstyle LV}}=\sqrt{\frac{A}{\pi\sigma_{\scriptscriptstyle LV}}}\left(\frac{\Delta\rho g\delta^2h_{\scriptscriptstyle LV}\rho_{\scriptscriptstyle V}}{2k_{\scriptscriptstyle V}\mu_{\scriptscriptstyle V}\Delta T}\right)
\end{equation}
\begin{equation}
k''=k\delta^2\sqrt{\frac{4\pi\sigma_{\scriptscriptstyle LV}}{A}}
\end{equation}
\begin{equation}
\omega'=\omega\left(\frac{h_{\scriptscriptstyle LV}\rho_{\scriptscriptstyle V}\delta^2}{k_{\scriptscriptstyle V}\Delta T}\right)
\end{equation}
This leads to the general, characteristic equation for the temporal growth rate $i\omega$ of the perturbation after eliminating $b_1'$ using eqn. \ref{eqn: tempeqnperturb} and \ref{eqn: delteqnperturb}:
\begin{multline}
\left(\frac{i\omega'}{8}+\frac{ik''\pi_{LB\sigma_{\scriptscriptstyle LV}}}{80}+\frac{3}{2Ja}\right) \left(i\omega'+ik''\frac{\pi_{LB\sigma_{\scriptscriptstyle LV}}}{4}+k''^4\pi_{LP}-2\pi_{LP}k''^2+3 \right)\\+ i\omega'\frac{c}{4}+ik''\frac{(1+c)\pi_{LB\sigma_{\scriptscriptstyle LV}}}{20}+k''^4\frac{3\pi_{LP}}{20}(\frac{7}{3}+c)-k''^2(\frac{7}{3}+c)\frac{3\pi_{LP}}{10}-\frac{3}{Ja}=0
\label{eqn: fulleqnnondim}
\end{multline}
The marginal state occurs when the real part $\text{Re}(i\omega)=0$, separating zones of stability ($\text{Re}(i\omega)<0$), where the perturbation amplitude decays in time, from regions of instability ($\text{Re}(i\omega)>0$), where the base state becomes unstable (Fig. \ref{fig: discrimja} \textbf{a}). Note that only three dimensionless numbers $Ja$, $\pi_{LP}$ and $\pi_{LB\sigma_{\scriptscriptstyle LV}}$ govern the stability of the perturbed solution. With the inclusion of van der Waals interactions, the buoyancy terms described by $\pi_{LB\sigma_{\scriptscriptstyle LV}}$ become negligible at nanoscale, as will be discussed shortly. 

This analysis incorporated time variation and convective transport in the energy equation. We can obtain a simpler expression for the stability problem by neglecting these two terms:

\begin{equation}
i\omega' = -\left(k''^4\pi_{LP}-2k''^2\pi_{LP}+1\right)-ik''\frac{\pi_{LB\sigma_{\scriptscriptstyle LV}}}{4}
\label{eqn: diffeqnnondim}
\end{equation}
where the buoyancy term $\pi_{LB\sigma_{\scriptscriptstyle LV}}$ is explicitly shown to give the dimensionless traveling wave velocity, signifying that the base flow acts only to convect the perturbation and does not affect its growth. The full derivation of eqn. \ref{eqn: diffeqnnondim} is provided in the Supporting Information. 

The diffusive expression (eqn. \ref{eqn: diffeqnnondim}) is a good estimate to the full stability equation (eqn. \ref{eqn: fulleqnnondim}) for small Jakob numbers, as demonstrated in Fig. \ref{fig: discrimja} \textbf{c}. Since the dimensionless parameters are calculated from the vapor properties at the superheated wall temperature, the Jakob number is small ($Ja\lessapprox\frac{1}{10}$) for the vapor phase of most fluids, implying that the thermal energy imparted by the heated solid is predominantly consumed through the latent heat of phase change rather than as sensible heat in raising the temperature of the vapor. Physically, dropping the time varying term in the energy equation implies that the heat conduction time scale is much longer (quasi-steady) than that for the perturbation growth in the phase change equation.

The Leidenfrost point corresponds to the lowest, critical $\pi_{LP, \;\text{crit}}$ on the marginal stability curve, below which the flow becomes unconditionally stable for all values of $k''$. For eqn. \ref{eqn: fulleqnnondim}, a good approximation for the critical $\pi_{LP}$ can be derived by noting that due to how we scaled the dimensionless parameters, $\pi_{LP, \;\text{crit}}$ occurs at $k''=1$. This leads to an algebraic equation for $\pi_{LP, \;\text{crit}}$  :
\begin{multline}
(\frac{Ja}{10}(\frac{7}{3}+c)+1)\pi_{LP, \;\text{crit}}-1)(1+\frac{Ja}{4}+\frac{Ja}{6}c-\frac{Ja}{12}\pi_{LP, \;\text{crit}})^2\\=(\frac{3\pi_{LB\sigma_{\scriptscriptstyle LV}}}{20})^2(\frac{Ja}{12})^2(1+\frac{Ja}{9}(2+c))(\frac{1}{3}+\frac{2c}{3}-\pi_{LP, \;\text{crit}})
\label{eqn: picritfull}
\end{multline}
Eqn. \ref{eqn: picritfull} was verified against a numerical solution to the full stability equation (eqn. \ref{eqn: fulleqnnondim}) with $\text{Re}(i\omega)=0$, and was found to give the same solution for $\pi_{LP, \;\text{crit}}$ up to machine precision for all parameter sets tested (Fig. \ref{fig: discrimja} \textbf{b}). For small Jakob numbers, Fig. \ref{fig: discrimja} \textbf{c} shows that the critical $\pi_{LP}$ can also be estimated from the diffusive expression (eqn. \ref{eqn: diffeqnnondim}).
\begin{equation}
\pi_{LP,\; \text{crit}}=1
\label{eqn: picritdiff}
\end{equation}
The implication of eqn. \ref{eqn: picritdiff} as a good estimate is threefold. First, it means that the boundary layer approximation is not required if film boiling is assumed to be diffusion dominated in both energy and momentum via lubrication theory. Secondly, the change in $\delta(x)$ along the streamwise direction is only captured in the base solution and becomes insignificant in the perturbation equations. Lastly, perturbation growth is independent of the base flow, which carries vapor generated at the liquid-vapor interface out of the local control volume. This explains why the LFP is not found to be strongly dependent on the configuration of the experimental set up. Any configuration eventually takes the vapor out of the film by some buoyancy driven force (even if it is horizontal or upside down - vapor eventually finds its way up). The strength of that driving force would indeed depend on configuration, but it does not matter for the perturbation solution.

\subsection{Stabilizing Terms}
The diffusive approximation to the critical $\pi_{LP}$ (eqn. \ref{eqn: picritdiff}) reveals that the main stabilizing terms are the liquid-vapor surface tension $\sigma_{\scriptscriptstyle LV}$, the evaporative phase change that replenishes local vapor mass $\frac{k_{\scriptscriptstyle V}\Delta T}{h_{\scriptscriptstyle LV}\rho_{\scriptscriptstyle V}}$, and the viscous shear $\mu$ that reduces mass transport away from any given point in the vapor field.  

The liquid-vapor surface tension acts as a restoring force against oscillatory modes imposed on the basic, locally-parallel solution. Positive curvature of the liquid-vapor interface with its center in the vapor region (curving into the liquid), induces high pressure locally with an adjacent low pressure zone due to the negative curvature of the continuous two phase interface. This creates a pressure gradient that attempts to restore the basic state by dampening all possible oscillatory frequencies. 

Similarly, a perturbed interface that bulges into the vapor steepens the thermal gradient in the vapor film, triggering an increase in the rate of evaporation locally that restores the base state and vice versa. Viscous shear is larger for smaller film thicknesses, therefore reducing mass transport away from a local bulge into the vapor domain and enhancing transport away from a bulge into the liquid field; this also acts to dampen perturbed modes. 

\subsection{van der Waals Interaction}
The heterogeneous Hamaker constant $A_{SVL}$ is used to characterize the van der Waals dispersion forces between an uncharged surface and an adjacent liquid separated by vacuum. It is incorporated into this analysis via the generalized pressure gradient (eqn: \ref{eqn: genP}). From the diffusive expression of the perturbation stability (eqn. \ref{eqn: diffeqnnondim}), it is shown that these dispersion forces between the liquid and solid substrate across the vapor film is the only destabilizing term for attractive interactions $A>0$. The film is unconditionally stable if the interaction is purely repulsive $A<0$. The former case holds in general for a liquid separated from a solid by a vacuum or an intermediate gas phase \cite{2002HamaVac, 2011Jacob}.

The relationship between the heterogeneous Hamaker constant and the contact angle $\theta$ of the substrate has been derived using Lifshitz theory \cite{1997Drummond, 1980houghwhite} :
\begin{equation}
1+\text{cos}(\theta)=\frac{A_{SVL}}{12\pi\sigma_{\scriptscriptstyle LV}H_{SVL}^2}
\label{eqn: thetaA}
\end{equation}
where $H_{SVL}$ is the equilibrium contact separation between the solid substrate (S) and the liquid (L) across vacuum (V) and takes on values in the order of magnitude of $1$ nm for most materials. Eqn. \ref{eqn: thetaA} can thus be used to account for the effect of surface wettability on the stability of the perturbed solution. As the van der Waals interaction only plays a significant role for film thicknesses that have reached the same order of magnitude as $H_{SVL}$, we approximate the ratio $\frac{H_{SVL}}{\delta}\approx 1$.

In this nanoscale regime, the neutral curve described by the full perturbation solution (eqn: \ref{eqn: fulleqnnondim}) is insensitive to $\pi_{LB\sigma_{\scriptscriptstyle LV}}$, which encapsulates the buoyancy force on the vapor film and is on the order of $1\mathrm{e}{-14}$. The diffusive expression of the perturbation equation (eqn. \ref{eqn: diffeqnnondim}) has an explicit dependence on $\pi_{LB\sigma_{\scriptscriptstyle LV}}$ only in the imaginary part of the temporal growth rate, such that the marginal state predicted is completely agnostic to changes in $\pi_{LB\sigma_{\scriptscriptstyle LV}}$. This implies that the stability criterion (eqn. \ref{eqn: picritfull} or \ref{eqn: picritdiff}) can be applied to capture the Leidenfrost point on plates of arbitrary orientation, as the direction and magnitude of the gravitational field does not play a significant role in the instability mechanism examined. The further implications of the nanoscale regime on this analysis is presented in the Supporting Information.

\section{Verification by experiment and simulation} 
To compare against experimental data, eqn. \ref{eqn: picritdiff} for the critical $\pi_{LP}$ can be rewritten in terms of material properties:
\begin{equation}
\frac{3}{(24\pi)^2}(\frac{1}{\delta})^4\frac{h_{\scriptscriptstyle LV}\rho_{\scriptscriptstyle V}\delta}{\sigma_{\scriptscriptstyle LV}}A^2\frac{1}{k_{\scriptscriptstyle V}\Delta T\mu_{\scriptscriptstyle V}}=1
\label{eqn: expandeddiff}
\end{equation}
Substituting in eqn. \ref{eqn: thetaA}, we obtain the corresponding expression for the LFP in terms of the intrinsic contact angle on the substrate. 
\begin{equation}
\frac{3}{4}\underbrace{\left(\frac{H_{SVL}}{\delta}\right)^4}_{\textstyle \approx 1\mathstrut}\underbrace{\left(\frac{h_{\scriptscriptstyle LV}\rho_{\scriptscriptstyle V}\delta}{\sigma_{\scriptscriptstyle LV}}\right)}_{\textstyle \pi_{2}\mathstrut}(1+\text{cos}(\theta))^2\underbrace{\frac{\sigma_{\scriptscriptstyle LV}^2}{k_{\scriptscriptstyle V}\Delta T\mu_{\scriptscriptstyle V}}}_{\textstyle \pi_{1}\mathstrut}=1
\label{eqn: expandedthetdiff}
\end{equation}
where we have defined two new dimensionless parameters that we will show to be significant:
\begin{align}
\pi_1&=\frac{\sigma_{\scriptscriptstyle LV}^2}{k_{\scriptscriptstyle V}\mu_{\scriptscriptstyle V}\Delta T} \\
\pi_2&=\frac{h_{\scriptscriptstyle LV}\rho_{\scriptscriptstyle V}\delta}{\sigma_{\scriptscriptstyle LV}}
\label{eqn: pi_2}
\end{align}

Eqn. \ref{eqn: expandedthetdiff} provides an explicit relationship between the intrinsic contact angle of a fluid on a substrate and the Leidenfrost point for the system. Since each fluid property ($k_{\scriptscriptstyle V}(T), \mu_{\scriptscriptstyle V}(T), \sigma_{\scriptscriptstyle LV}(T)$, etc.) is calculated at the superheated wall temperature, the left hand side of eqn. \ref{eqn: expandedthetdiff} is in general a nonlinear function of temperature. The temperature at which eqn. \ref{eqn: expandedthetdiff} is satisfied corresponds to the predicted LFP; this can be found numerically with the temperature and pressure dependent fluid properties available from databases like NIST and tabulations from literature \cite{2012NIST, Cheric, 2006Transport}. 
\subsection{Surface dependence of the LFP}
\begin{figure}[]
\centering
\includegraphics[width=8.5cm]{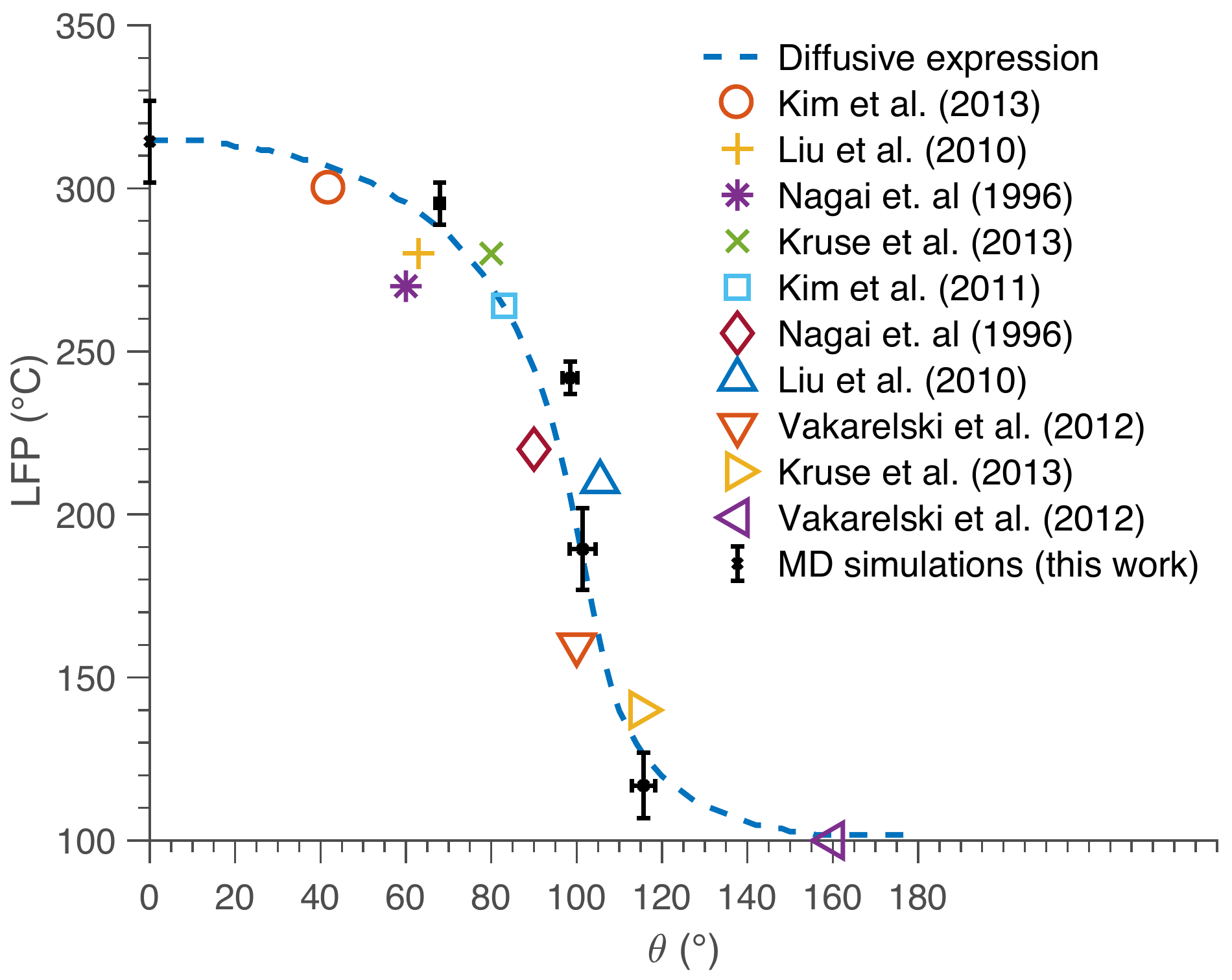}
\caption{The Leidenfrost temperature vs the contact angle for water, from experiments  \cite{2011WaterLFP, 2013Kruse, 2012Stabiliz, 2013Nanotubes, 2010macroflat}, the diffusive prediction of the LFP (eqn. \ref{eqn: expandedthetdiff}) and molecular dynamics simulations from this work. The equilibrium separation $H_{SVL}$ and its variation associated with changes in the contact angle $\dv{H_{SVL}}{\theta}$ can be estimated from experimental data \cite{2000waterHama, 1997Drummond}. Typical errors in the LFP and the contact angle measured from experiment are around $5\degree\text{C}$ and $2\degree$ respectively, though many sources do not explicitly report an error value for either quantity. The data point corresponding to a contact angle of $160\degree$ from Vakarelski et al. corresponds to the only surface which was textured with nanoparticles to achieve superhydrophobicity; the other data points correspond to flat surfaces without deliberate texturing.}

\label{fig: thetawater}
\end{figure}

The LFP for water has been demonstrated to vary dramatically with changes in the liquid wettability of the solid surface \cite{2013Nanotubes, 2013Kruse}. Fig. \ref{fig: thetawater} shows that the diffusive prediction of the LFP (eqn. \ref{eqn: expandedthetdiff}) accurately captures the relationship between the LFP and the contact angle as delineated by experiments \cite{2011WaterLFP, 2013Kruse, 2012Stabiliz, 2013Nanotubes, 2010macroflat}. Physically, larger contact angles indicate a hydrophobic substrate, which exhibits less attractive van der Waals interaction with the bulk liquid and presents a smaller destabilizing effect to the vapor film; the LFP thus decreases to near the boiling point. Without considering van der Waals interaction between the liquid and substrate surfaces, such a relationship cannot be explained or predicted from first principles.

Further evidence of the significant role played by van der Waals forces in governing the LFP arises from X-ray imaging of the vapor film collapse \cite{2019Jones}. Images spanning the film lifespan between formation and collapse showed that film collapse on the macroscopic level is preceded by submicron length scale vapor film thicknesses where the bulk liquid appears to wet the substrate. Although instabilities on the micron scale and above perturb the liquid-vapor interface and induce frequent local contact between the liquid and solid, only when the vapor film becomes unstable on the smallest length scales where van der Waals interactions dominate will the film completely collapse. 
\begin{figure}[]
\centering
\includegraphics[width=8.5cm]{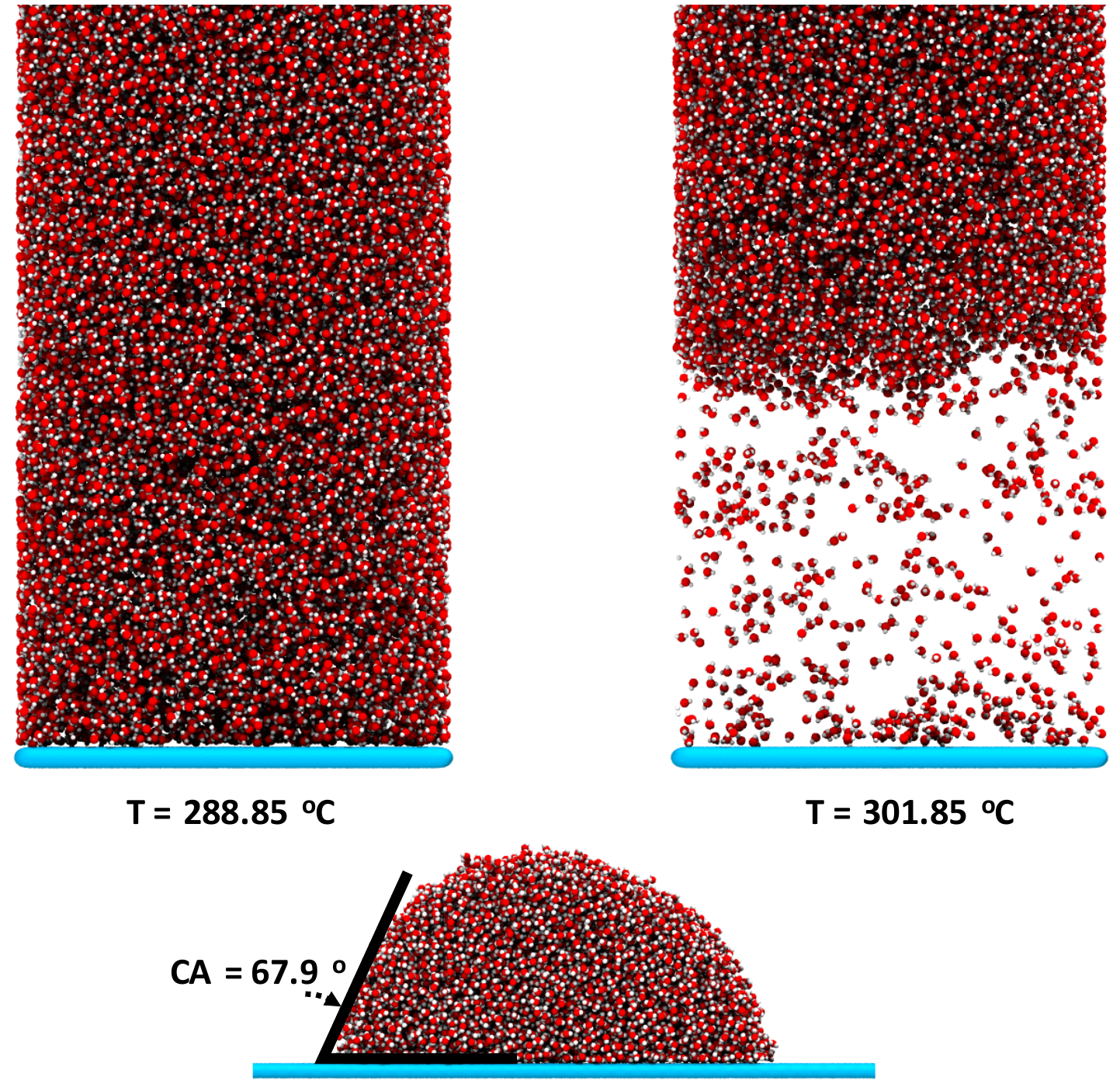}
\caption{Molecular dynamics simulation of vapor film formation adjacent to a heated surface. The system is pressurized at 1 atm.}

\label{fig: MDsims}
\end{figure}

Lastly, we note that our theoretical analysis predicts that the main effects governing the LFP operate in the nanoscale regime, which is accessible by molecular dynamics (MD). Figure \ref{fig: MDsims} shows one of the boiling heat transfer simulations performed using LAMMPS \cite{1995LAMMPS} software to numerically determine the LFP and the corresponding intrinsic contact angle of the substrate. Details of the MD implementations are provided in the Supporting Information. Note that a vapor film forms when the liquid water adjacent to the bottom plate is heated above the LFP, whereas liquid contact with the solid surface is preserved below the LFP due to the attractive heterogeneous van der Waals interactions. The relationship between the contact angle and the Leidenfrost point of the SPC/E (extended simple point charge) water model is in good agreement with the diffusive prediction of the LFP (eqn. \ref{eqn: expandedthetdiff}). These simulations show that vapor film stability is ultimately determined at the proposed nanometric length scale where fluid-surface van der Waals interactions cannot be discounted and where the effect of gravity driven instabilities is nonexistent.
\subsection{Fluid dependence of the LFP}

For most experimentally available data on the Leidenfrost point, the contact angle of the fluid on the substrate material is low, around $\theta\approx20\degree$. Nonetheless, the Hamaker constant must be found to determine the equilibrium separation $H_{SVL}$. Although the assumption $\frac{H_{SVL}}{\delta}\approx 1$ is made, the base film thickness $\delta$ still needs to be incorporated into our instability expression via $\pi_2$ (eqn: \ref{eqn: pi_2}). We can find the homogeneous Hamaker constant of the fluid (acentone, ethnanol, benzene, etc.) and the substrate (gold, aluminum, copper), and take the geometric mean to obtain the heterogeneous value \cite{2011Jacob, 2012Leite}. From the relationship between the surface energy and homogeneous Hamaker constant, we can obtain the homogeneous contact separations via: 
\begin{equation}
\sigma_{\scriptscriptstyle LV}=\frac{A_{LVL}}{24\pi H_{LVL}^2}
\end{equation}
\begin{equation}
\sigma_{SV}=\frac{A_{SVS}}{24\pi H_{SVS}^2}
\end{equation}

Figure \ref{fig: alkanesseparation} shows that it is possible to determine either the heterogeneous Hamaker constant given the Leidenfrost point for a fluid on a solid substrate, or vice versa with knowledge of the homogeneous Hamaker constants of both species. 

In general, experimental data on the homogeneous Hamaker constants may not be available for a fluid or substrate of interest. Here, we note an avenue for simplification: it is observed that the dimensionless quantity $\pi_2=\frac{h_{\scriptscriptstyle LV}\rho_{\scriptscriptstyle V}\delta}{\sigma_{\scriptscriptstyle LV}}$ in the diffusive expression is around $0.06$ for most fluids at their respective Leidenfrost temperatures. This suggests that there exists a functional dependence $H_{SVL}=F(\frac{h_{\scriptscriptstyle LV}\rho_{\scriptscriptstyle V}}{\sigma_{\scriptscriptstyle LV}})$. Additionally, most experimental setups in the film boiling regime feature fluids that wet the surface in contact, such that their intrinsic contact angle are small ($\theta \approx 20\degree$) \cite{2006Transport}. 
\begin{figure}[]
\centering
\includegraphics[width=8.5cm]{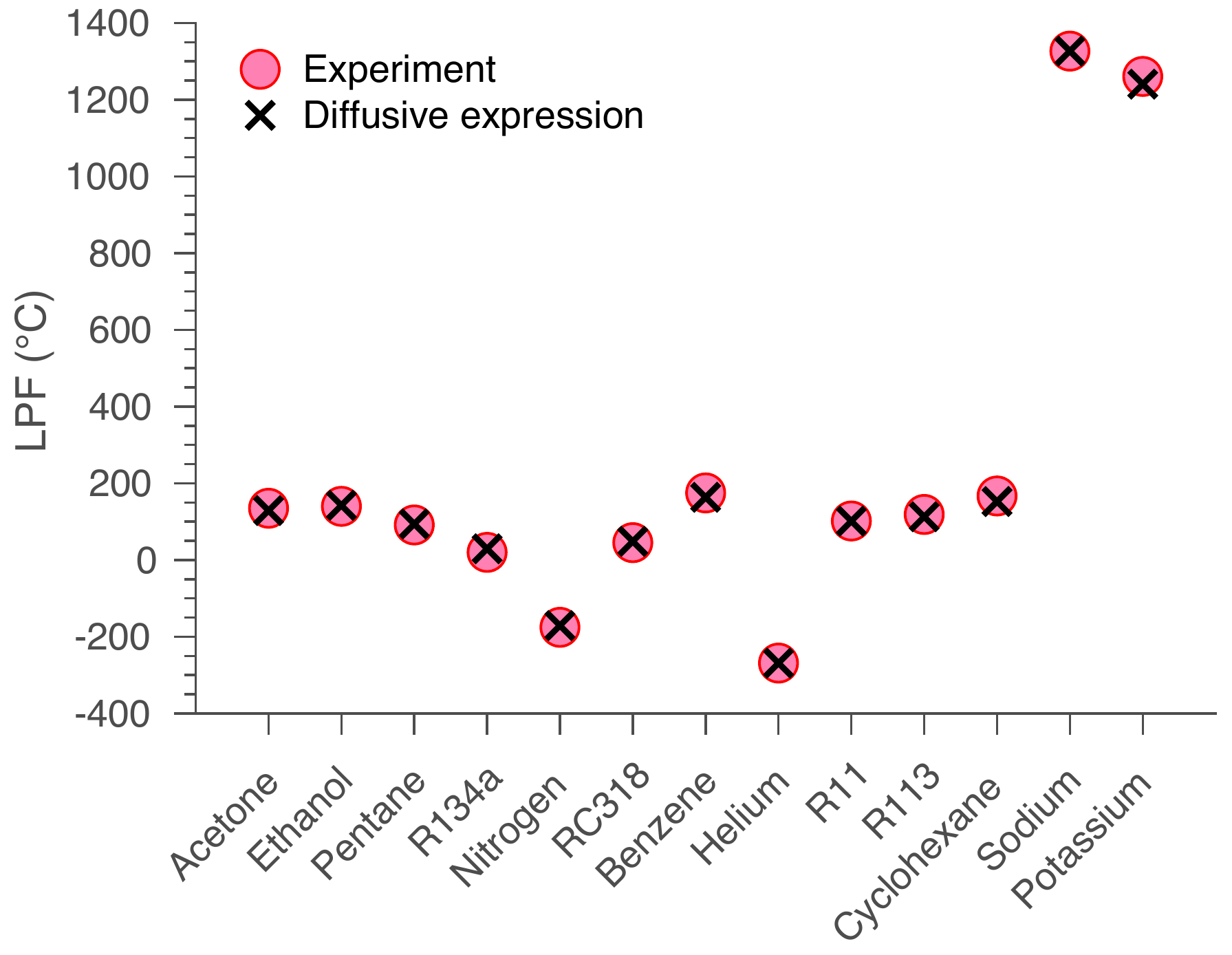}
\caption{The dimensionless criterion $\pi_1=6$ as a low contact angle approximation from the diffusive expression (eqn. \ref{eqn: expandedthetdiff}) captures the LFP data from experiment to within 5\% error. The dimensionless number $\pi_1$ encapsulates the stabilizing effects of evaporative phase change (vapor mass generation), surface tension and viscous transport, while the critical value at which the LFP occurs describes the destabilizing role of the van der Waals interaction for low contact angle fluids. Experimental LFP and fluid property data are available for acetone, ethanol, pentane, R134a, nitrogen, RC318, benzene, helium, R11, R113, liquid sodium and liquid potassium \cite{1999Acetone, 1986Alcohols, 1982Hydrocarbons, 2006R134a, 2018Nitrogen, 1985RC318, 1973minleid, 2012NIST, Cheric, 2006Transport,1996Nagai}
}
\label{fig: pi1=6}
\end{figure}
From the diffusive expression (eqn. \ref{eqn: expandedthetdiff}) valid for low Jakob numbers, the above approximations leads to a simplified, dimensionless prediction to the Leidenfrost point for fluids/substrate systems with low, intrinsic contact angles:
\begin{equation} 
 \pi_1=\frac{\sigma_{\scriptscriptstyle LV}^2}{k_{\scriptscriptstyle V}\Delta T\mu_{\scriptscriptstyle V}}\approx6
 \label{eqn. pi16}
 \end{equation}
This dimensionless quantity also arises by application of the Buckingham's Pi Theorem to the system, as discussed in the Supporting Information. Fig. \ref{fig: pi1=6} shows that the temperature at which this equality is satisfied captures the experimental data on the LFP for a variety of different fluids, including cryogens and liquid metals. The single dimensionless number describes the terms that stabilize the vapor film, including surface tension, phase change and viscous transport, while the critical value corresponding to the LFP denotes the destabilizing effect of attractive van der Waals interaction between the bulk liquid and solid substrate. Larger values of $\pi_1$ above the critical imply the system is the film boiling regime, since the stabilizing terms dominate.
 
 \subsection{Pressure dependence of the LFP}
 \begin{figure}[]
\centering
\includegraphics[width=8.5cm]{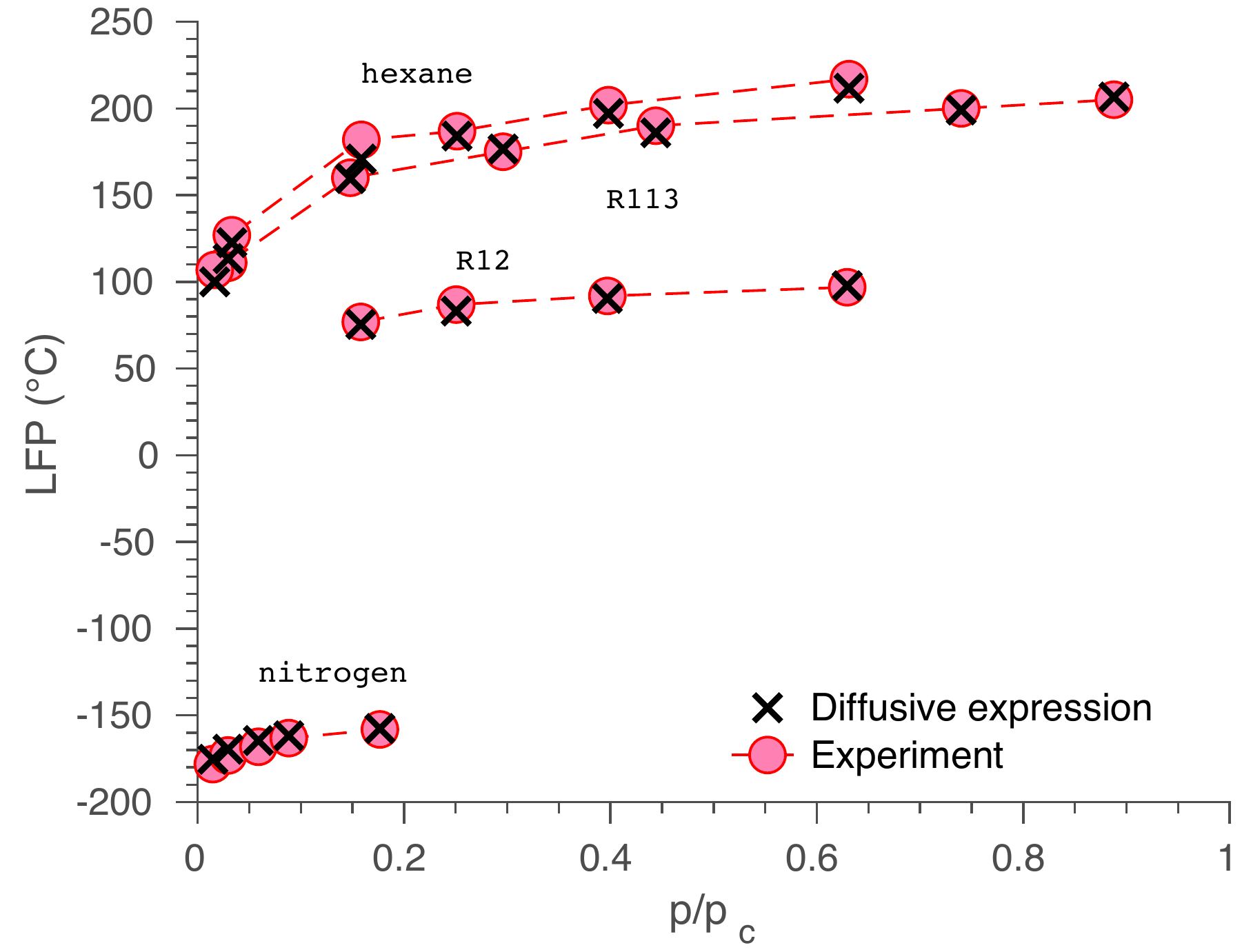}
\caption{The LFP from experimental data of R113, R12, hexane and nitrogen at different pressures (o's) \cite{1987hexanepress, 1978Pressure, 1990Sakurai} vs the predictive ability of the diffusive expression (eqn. \ref{eqn: expandedthetdiff}).}

\label{fig: pi1=6pressure}
\end{figure}
Experimental work has shown that the LFP depends on the ambient pressure applied, such that the Leidenfrost temperature gradually increases from near the boiling point towards the critical point of the fluid \cite{1978Pressure}. Figure \ref{fig: pi1=6pressure} demonstrates that the diffusive expression (eqn. \ref{eqn: expandedthetdiff}) captures the LFP of various fluids for both subatmospheric and superatmospheric pressures up to the critical point.
\section{Conclusion}
The dynamic stability of a vapor film on a heated vertical wall under the effects of gravity were considered. The only possible instability at nanoscale was driven by attractive van der Waals interaction between the bulk liquid and the substrate, which could be stabilized by the liquid-vapor surface tension, evaporative phase change and viscous transport. The marginal or neutral state can be found analytically (eqn. \ref{eqn: picritfull}) for the most general case, or simplified for small Jakob number flows to a diffusive approximation (eqn. \ref{eqn: picritdiff}). The resulting theoretical solution for the LFP captures the variation of experimental data with surface wettability, fluid properties and pressure.

A single, dimensionless number $\pi_1$  is found to encapsulate the physical instability mechanism of the Leidenfrost phenomenon for wetting fluids. The value of $\pi_1$ with respect to the critical denotes regimes in which the vapor film is stable or unstable, providing a useful characterization of both the thermodynamic state and the physical means by which transition to the pool boiling regime occurs.

This insight into the nanoscale mechanisms inducing the transition from film to nucleate boiling enables control of the phase adjacent to the surface \cite{2019phasecontrol}. It would be of interest to extend the instability mechanism towards surface roughness, which experiment has shown to effect dramatic changes in the LFP beyond what can be explained by variation in surface wettability \cite{2013Varanasi, 2013Kruse, 2018drops}. In addition, a theoretical treatment of the Nukiyama temperature corresponding to the critical heat flux may reveal the mechanism underlying transition boiling and provide a comprehensive understanding of the entire boiling curve under a unified, physical framework.

\section*{Author contributions}
N.A.P. and T.Y.Z. conceived and planned the research, performed the analyses, and wrote the manuscript.
\section*{Competing interests}
The authors have no competing financial interests or other interests that might be perceived to influence the results and/or discussion reported in this paper.
\section*{Corresponding author}
To whom correspondence should be addressed. E-mail: n-patankar@northwestern.edu

\beginsupplement
\section*{Supporting Information (SI)}
\subsection*{Buckingham's Pi Theorem}
In general, the thermodynamic variables that affect the physical onset of film boiling include the thermal conductivity of the vapor ($k_V$), temperature ($T$), viscosity of the vapor ($\mu_V$), liquid-vapor surface tension ($\sigma_{LV}$), the specific latent heat of vaporization ($h_{LV}$), the liquid density $\rho_L$, the vapor pressure $P_V$, and the gravitational constant $g$ \cite{1961Berenson, 1973minleid, 1992Carey}. This leads to four dimensions, namely mass $M$, length $L$, time $\tau$ and temperature $\phi$. Taking the product $k_V(\Delta T)=k_V(T_w-T_s)$ as a single variable, the number of dimensions can be reduced to three, forming the dimensional matrix \cite{2010Kundu}:

\begin{table}[h]
\centering
\begin{tabular}{c | c c c c c c c}
 & $\sigma_{LV}$ & $\mu_V$ & $k_V(T_w-T_s)$ & $h_{LV}$ & $\rho_L$ & $g$ & $P_V$\\ \hline
$M$ & 1                     & 1                  & 1  & 0          & 1      & 0  &   1  \\
$L$ & 0                     & -1                 & 1  & 2            &  -3       & 1 &  -1\\
$\tau$ & -2                    & -1                 & -3 & -2             & 0        &  -2 & -2\\ 
\end{tabular}
\end{table}
The repeating variables are chosen to be $\sigma_{LV}$, $k_V(T_w-T_s)$ and $\rho_L$, which span the space of $M$, $L$, and $\tau$. Given seven variables and a rank three matrix, four dimensionless numbers are expected: $\Pi_1=\frac{\sigma^2}{k_V(T_w-T_s)\mu_V}$, $\Pi_2=\frac{\sigma_{LV}^2 h_{LV}}{k_V^2(T_w-T_s)^2}$, $\Pi_3=\frac{\sigma_{LV}^5 g}{k_V^4(T_w-T_s)^4\rho_L}$, $\Pi_4=\frac{\sigma^2P_V}{k_V^2(T_w-T_s)^2\rho_L}$. We note immediately that for the dimensionless numbers $\Pi_1$ and $\Pi_2$, the repeating variable $\rho_L$ is absent. This suggests that density or mass is not relevant to the subset of variables $k_V(T_w-T_s)$, $\sigma_{LV}$, $h_{LV}$ and $\mu_V$; the dimensionless combination of these variables may be responsible for the static equilibrium behavior of a fluid in the different boiling regimes.   

Thus, to reduce the number of independent variables describing the LFP, the gravity, pressure and density of either phase are postponed for consideration in the present analysis. Prior work has shown that accounting for gravity and the density ratio between liquid and vapor accurately captures the length scales of the droplet and vapor film as well as their evolution in time specific to the Leidenfrost regime \cite{2003Quere}. The present work focuses instead on the characteristics of a fluid that determines its LFP and distinguishes film boiling from nucleate or transition boiling. 

By reducing the variable space, we obtain the dimensional matrix:

\begin{table}[]
\centering
\begin{tabular}{c | c c c c}
 & $\sigma_{LV}$ & $\mu_V$ & $k_V(T_w-T_s)$ & $h_{LV}$ \\ \hline
$M$ & 1                     & 1                  & 1  & 0                       \\
$L$ & 0                     & -1                 & 1  & 2                       \\
$\tau$ & -2                    & -1                 & -3 & -2                      \\ 
\end{tabular}
\end{table}
The rank of the matrix is two as the rows are linearly dependent; since density, pressure and gravity are neglected, mass is not relevant to the problem. With four variables and a rank two matrix, two dimensionless numbers are expected from the analysis. The repeating variables are chosen to be $\sigma_{LV}$ and $k_V(T_w-T_s)$, which yield the dimensionless numbers:
\begin{equation}
\Pi_1=\frac{\sigma^2}{k_V(T_w-T_s)\mu_V}=\pi_1
\end{equation}
\begin{equation}
\Pi_2=\frac{\sigma^2\Delta H}{k^2_V(T_w-T_s)^2}
\end{equation}
The dimensionless number $\pi_1$ represents the ratio of the liquid-vapor surface tension to the viscosity, thermal conductivity and temperature of the vapor phase. The LFP is posited to increase for fluids with larger liquid-vapor surface tension based on the model for the maximum superheat limit \cite{1992Carey} and the Taylor type instability \cite{1961Berenson}. The same correlation between surface tension and the LFP has been demonstrated experimentally across a broad range of fluids \cite{1973minleid}. From the superheat limit viewpoint, the relationship arises since a lower liquid-vapor surface tension decreases the energy barrier and increases the rate of vapor nucleation, which reduces the critical temperature necessary to form a vapor film. 

From the Taylor instability perspective, the role of surface tension appears contradictory. Larger surface tension is theorized to stabilize the liquid-vapor interface by suppressing low wavelength perturbations and reducing the growth of high wavelength disturbances \cite{1961Berenson}. It is therefore expected that higher surface tension would lead to a lower LFP by promoting the formation of a stable vapor film. However, the predicted expression for the Leidenfrost temperature originating from Taylor instability theory proposes that the LFP should increase with the surface tension, based on a fitted expression for bubble radii \cite{1961Berenson}. While accurate for n-pentanes and carbon tetrachloride, it has been shown experimentally that the LFP predicted from the instability analysis is less accurate for water, cryogenic fluids and liquid metals \cite{1973minleid}.      

The experimentally corroborated relationship between liquid-vapor surface tension and the LFP of a fluid means that the Leidenfrost regime should occur for small values of the dimensionless number $\pi_1$, corresponding to low surface tension, high vapor viscosity, high vapor thermal conductivity and high temperature; nucleate or transition boiling occurs for large values of $\pi_1$. For consistency, this suggests that higher vapor thermal conductivity should trigger the Leidenfrost phenomenon by conveying more heat to the liquid-vapor interface and boosting the rate of vaporization. A higher vapor viscosity should also induce film boiling by stabilizing the vapor layer from perturbations and penalizing the escape of the vapor phase from the film.

Physically, $\pi_1$ compares two velocities $\frac{k_V(T_w-T_s)}{\sigma}$ and $\frac{\sigma}{\mu_V}$; the units of these quantities are $\frac{L}{\tau}$. The former $\frac{k_V(T_w-T_s)}{\sigma_{LV}}$ gives a velocity scale of the vapor phase in the direction normal to the liquid vapor interface, which is approximately parallel to the heated surface. The latter $\frac{\sigma_{LV}}{\mu_V}$ gives a velocity scale of the vapor phase in the direction parallel to the liquid vapor interface. If the normal component of the velocity $\frac{k_V(T_w-T_s)}{\sigma_{LV}}$ exceeds the tangential component $\frac{\sigma_{LV}}{\mu_V}$, a film of vapor can be sustained beneath the droplet. The transition to the Leidenfrost regime therefore occurs at $\pi_1$ around order one. 

The second dimensionless number $\Pi_2$ represents the ratio of the liquid-vapor surface tension and the latent heat of vaporization to the thermal conductivity and temperature of the vapor phase. The Leidenfrost regime should occur for small values of $\Pi_2$, corresponding to low surface tension, high vapor thermal conductivity, high temperature and low latent heat of vaporization. With all else constant, a lower latent heat can trigger the Leidenfrost effect by increasing the rate of vaporization to sustain a stable film \cite{1992Carey}. 

Physically, $\Pi_2$ compares the specific energies $h_{LV}$ and $\frac{k^2_V(T_w-T_s)^2}{\sigma_{LV}^2}$. The energy scale $h_{LV}$ gives the specific latent heat needed to sustain vaporization at steady state. The latter $\frac{k^2_V(T_w-T_s)^2}{\sigma_{LV}^2}$ represents the energy scale for heat conducted to the liquid-vapor interface. If the conducted heat $\frac{k^2_V(T_w-T_s)^2}{\sigma_{LV}^2}$ that reaches the interface exceeds the specific latent heat $h_{LV}$, a vapor film can be sustained beneath the liquid. The transition to the Leidenfrost regime may thus occur for $\Pi_2$ around order one. 

\subsection*{Molecular dynamics}
Molecular dynamics simulations were implemented using LAMMPS \cite{1995LAMMPS} software. Liquid water was initially equilibrated at saturation temperature in the canonical ensemble with constant pressure imposed by a piston constrained to move only in the direction orthogonal to the bottom surface \cite{2016Jones}. The solid plate was constructed using a graphene sheet, and the interaction between the solid substrate and the SPC/E water molecules was governed by the 6-12 Lennard Jones pair potential. After equilibration, the liquid adjacent to the bottom surface was heated to a target temperature around the LFP, whereas the liquid adjacent to the piston was maintained at constant, saturation temperature to simulate nonequilibrium heat transfer conditions \cite{2014heatconduction}. 
\bibliographystyle{unsrt}
\subsection*{Relationship between the Hamaker constant and the equilibrium separation}
\begin{figure}[]
\centering
\includegraphics[width=8.5cm]{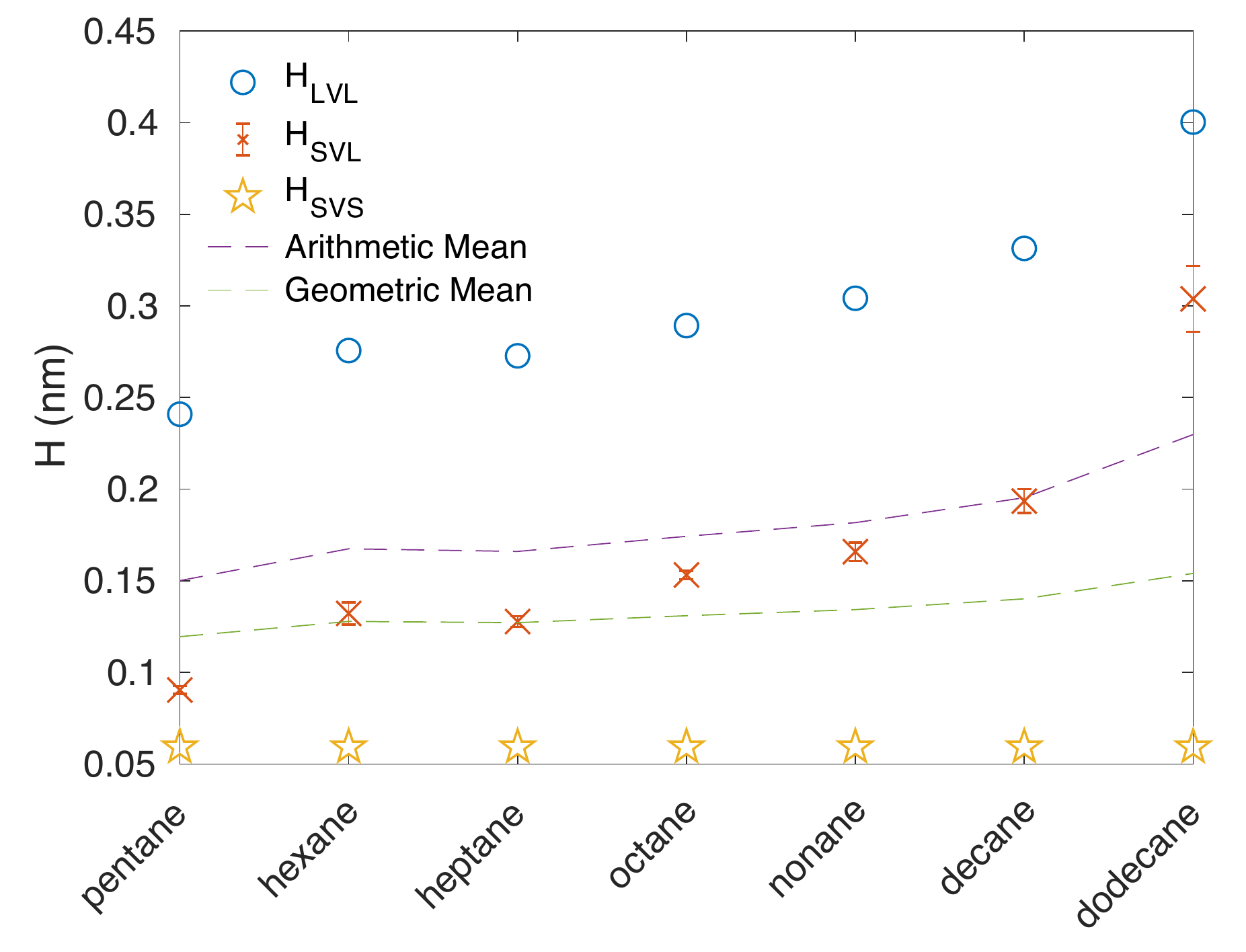}
\caption{The equilibrium contact surface separation for the fluid (alkanes), surface (gold) and the heterogeneous value corresponding to the Leidenfrost point. Note that the equilibrium separation $H_{LVL}$ is reported at the respective Leidenfrost temperatures of each alkane species, whereas Drummond et. al. listed $H_{LVL}$ at the same temperature. The temperature dependence of $H_{SVS}$ is assumed to be small over the range of temperatures corresponding to the LFP of the alkane series \cite{2001Ederth, 2018goldca}, which is much lower than the melting point of gold.}
\label{fig: alkanesseparation}
\end{figure}

The heterogeneous contact separation have been estimated as the arithmetic or geometric mean of the homogeneous values \cite{1997Drummond}, although these means may not provide a good approximation for the actual $H_{SVL}$ in general. Nonetheless, for our theory to be physically consistent, the heterogeneous contact separation corresponding to the experimental Leidenfrost temperatures must be between the two bounding homogeneous values. Figure \ref{fig: alkanesseparation} shows that for the alkane family, this condition $H_{SVS}<\delta\approx H_{PVL}<H_{LVL}$ is satisfied, and the arithmetic and geometric means provide a reasonable estimate to the actual heterogeneous value. 

Drummond has shown that as the chain length of the alkane species increases, the contact angle increases correspondingly. This suggests that longer chain alkanes in the liquid phase have unfavorable energetic interactions with a given substrate (greater liquid-solid surface energy $\sigma_{SL}$) compared to small chain alkanes on the same solid material. The heterogeneous, equilibrium distance $H_{PVL}$ therefore tends to increase with the straight chain length of the alkane species, moving from near the solid separation $H_{SVS}$ toward the liquid value $H_{LVL}$. 

\subsection{Derivation of the diffusive equation}
We begin with the general formulation of the vertical film boiling setup, taking into account only the diffusive terms in the steady momentum and energy conservation equations:
\begin{equation}
\pdv{\bar{u}_V}{x}+\pdv{\bar{v}_V}{y}=0
\label{eqn: d mass conservation}
\end{equation}
\begin{equation}
0=-\frac{1}{\rho_V}\pdv{\bar{\Phi}}{x}+\frac{f_d}{\rho_V}+\frac{\mu_V}{\rho_V}\pdv[2]{\bar{u}_V}{y}
\label{eqn: d mom conservation}
\end{equation}
\begin{equation}
0=\frac{k_V}{\rho_Vc_{p,V}}\pdv[2]{\bar{\Theta}_V}{y}
\label{eqn: d energy conservation}
\end{equation}
where $f_d$ is a buoyancy driven force dependent on the system configuration. As before, the boundary conditions at the superheated wall and the liquid-vapor interface at $\bar{\delta}(x)$ are given by:
 \begin{equation}
\text{at} \; y=0, \quad \bar{u}_V=\bar{v}_V=0\;, \quad \bar{\Theta}_V=1
\label{eqn: d wallbcs}
\end{equation}
 \begin{equation}
\text{at}\; y=\bar{\delta}, \quad \bar{u}_V+\bar{v}_V\dv{\bar{\delta}}{x}=0\;, \quad \bar{\Theta}_V=0
\label{eqn: d intbcs}
\end{equation}
while the phase change equation at the interface is written as:
\begin{equation}
\rho_V\pdv{\bar{\delta}}{t}+\pdv{}{x}\int_0^{\bar{\delta}}(\rho_V\bar{u}_Vdy)=-\frac{k_V}{h_{LV}}\pdv{\bar{\Theta}_V}{y}|_{y=\bar{\delta}}
\label{eqn: d basedelta}
\end{equation}
The base solution for the velocity field as found using an integral expansion is identical to our prior solution to the boundary layer equations:
\begin{equation}
u=a_0+a_1\eta+a_2\eta^2=\frac{f_d\delta^2}{2\mu_V}(\eta-\eta^2)
\label{eqn: d baseu}
\end{equation}
The temperature field in the base solution is linear, as expected from a diffusive approximation:
\begin{equation}
\Theta=b_0+b_1\eta+b_2\eta^2=1-\eta
\label{eqn: d baseT}
\end{equation}
The corresponding perturbed solutions are:
\begin{equation}
u'=a'_0+a'_1\eta+a'_2\eta^2=(\frac{f_d\delta\delta'}{2\mu_V}-\frac{\pdv{\Phi'}{x}\delta^2}{2\mu_V})\eta+\frac{\pdv{\Phi'}{x}\delta^2}{2\mu_V}\eta^2
\label{eqn: d puT}
\end{equation}
\begin{equation}
\Theta'=b'_0+b'_1\eta+b'_2\eta^2=\frac{\delta'}{\delta}\eta^2
\label{eqn: d puT2}
\end{equation}
The variation of the film thickness in the streamwise direction is found from eqn. \ref{eqn: d basedelta} after assuming time invariance in the base state: 
\begin{equation}
\delta=\left(\frac{16k_V\Delta T\mu_V}{\rho_Vh_{LV}f_d} x \right)^{\frac{1}{4}}
\label{eqn: d delta}
\end{equation}
Note that compared with the solution for the film thickness obtained from the boundary layer equations, eqn. \ref{eqn: d deltalin} is missing the term $(1-\frac{1}{3}\frac{c_{PV}\Delta T}{h_{LV}})^(1/4)$. This is expected given only conduction is consumed in the thin film limit.

Linearizing the phase change conservation condition (eqn: \ref{eqn: d basedelta}) at the interface we obtain:
\begin{equation}
\rho_V\pdv{\delta'}{t}+\pdv{}{x}\int_0^\delta\rho_Vu'_Vdy+\pdv{}{x}\int_0^{\delta+\delta'}\rho_Vu_Vdy=-\frac{k\Delta T}{h_{LV}}\left(\delta'\pdv[2]{\Theta}{y}|_\delta+\pdv{\Theta'}{y}|_{\delta}\right)
\label{eqn: d deltalin}
\end{equation}
The change in $\delta'$ is obtained after substituting in the expressions for $u'$ and $\Theta'$ (eqn: \ref{eqn: d puT} and \ref{eqn: d puT2} respectively): 
\begin{equation}
\rho_V\pdv{\delta'}{t}+\frac{\rho_Vf_d \delta^2}{4\mu_V}\pdv{\delta'}{x}+\frac{\rho_V\delta^3\sigma}{12\mu_V}\pdv[4]{\delta'}{x}+\frac{\rho_VA}{24\pi\mu)V\delta}\pdv[2]{\delta'}{x}+\frac{k_V\Delta T}{h_{LV}\delta^2}\delta'=0
\label{eqn: d deltalin}
\end{equation}
Expressing in terms of normal modes:
\begin{equation}
\delta'=\delta'_a\text{exp}(i(kx+\omega t))
\end{equation}
where for brevity we take $\delta'_a\rightarrow\delta'$. This gives us:
\begin{equation}
i\omega=-\frac{k_V\Delta T}{h_{LV}\delta^2\rho_V}-\frac{\delta^3\sigma}{12\mu_V}k^4+\frac{A}{24\pi\mu_V\delta}k^2-i\frac{f_d\delta^2}{4\mu_V}k
\end{equation}
After substituting in the set of dimensionless numbers introduced previous, we recover the diffusive expression (eqn. \ref{eqn: diffeqnnondim}). 

\subsection{Nanoscale implications}
Given that the phenomenon is localized in the nanoscale regime, it is important to discuss the applicability of the Navier-Stokes equations in the context of O(1) in the Knudsen number (Kn). Hadjiconstantinou showed that the second order Knudsen layer correction is qualitatively robust well beyond $Kn\approx0.4$, such that the underlying Navier-Stokes constitutive laws captures the behavior of arbitrary flows in spite of superimposed kinetic corrections in the flow field \cite{2006Knudsen}. 

Here, the base solutions for velocity (eqn. \ref{eqn: d baseu}) and temperature (eqn. \ref{eqn: d baseT}) for the diffusive equation above are represented by the lowest order polynomials possible, given the buoyancy force in the momentum equation and no heat source in the energy equation. An examination of the system from a time and spatially averaged statistical standpoint would give us similar profiles with an effective slip and temperature jump at the interfaces, giving rise to effective viscosities and thermal conductivities that converge to the continuum values as the film thickness is increased.

It would be of great interest and rigor to frame the film stability problem in the context of a linearized Boltzmann analysis to accommodate larger Kn flows; however, the excellent agreement between the present, simplified model with experiment suggests the dominant effects governing the Leidenfrost film stability is captured despite the absence of statistical analyses describing intermolecular collisions. Additionally, the granularity of linear Boltzmann equation may obscure the effect of surface tension and van der Waals interactions, which have been shown to be important factors in pinpointing the LFP.

\newpage

\end{document}